\documentclass[12pt]{article}
\usepackage{natbib}
\usepackage{color}
\usepackage{latexsym}
\usepackage{amsthm}
\usepackage{multirow}
\usepackage{float}
\usepackage{wrapfig}
\usepackage[latin1]{inputenc}
\usepackage[T1]{fontenc}
\usepackage{amsmath,amsfonts,amssymb,bm,dsfont}
\usepackage{pslatex}
\usepackage{graphicx,color}
\usepackage{fancyhdr}
\usepackage{multicol}
\usepackage{thumbpdf,lmodern}

\makeatletter
\renewcommand\@biblabel[1]{}

\makeatother
\usepackage[displaymath]{lineno}
\usepackage[latin1]{inputenc}
\usepackage[T1]{fontenc}
\usepackage{graphicx}
\usepackage{times}
\usepackage{amsmath,amsfonts,amssymb,bm}
\usepackage[english]{babel}
\usepackage{amsbsy}

\usepackage{subfigure}
\usepackage{multirow}
\usepackage{multicol}
\theoremstyle{definition}

 \newcommand*\patchAmsMathEnvironmentForLineno[1]{%
  \expandafter\let\csname old#1\expandafter\endcsname\csname #1\endcsname
  \expandafter\let\csname oldend#1\expandafter\endcsname\csname end#1\endcsname
  \renewenvironment{#1}%
     {\linenomath\csname old#1\endcsname}%
     {\csname oldend#1\endcsname\endlinenomath}}%
\newcommand*\patchBothAmsMathEnvironmentsForLineno[1]{%
  \patchAmsMathEnvironmentForLineno{#1}%
  \patchAmsMathEnvironmentForLineno{#1*}}%
\AtBeginDocument{%
\patchBothAmsMathEnvironmentsForLineno{equation}%
\patchBothAmsMathEnvironmentsForLineno{align}%
\patchBothAmsMathEnvironmentsForLineno{flalign}%
\patchBothAmsMathEnvironmentsForLineno{alignat}%
\patchBothAmsMathEnvironmentsForLineno{gather}%
\patchBothAmsMathEnvironmentsForLineno{multline}%
}

\begin{document}

\begin{center}

\Large {\bf SpatEntropy: Spatial Entropy Measures in \texttt{R}}\par

\bigskip

\normalsize{Linda Altieri, Daniela Cocchi, Giulia Roli}\\ \small{Department of Statistical Sciences, University of Bologna, via Belle Arti, 41, 40126, Bologna, Italy.} \par

\bigskip

%

\end{center}

\begin{quotation}
\noindent {\it Abstract:}
This article illustrates how to measure the heterogeneity of spatial data presenting a finite number of categories via computation of spatial entropy. The \texttt{R} package \texttt{SpatEntropy} contains functions for the computation of entropy and spatial entropy measures. The extension to spatial entropy measures is a unique feature of \texttt{SpatEntropy}. In addition to the traditional version of Shannon's entropy, the package includes Batty's spatial entropy, O'Neill's entropy, Li and Reynolds' contagion index, Karlstrom and Ceccato's entropy, Leibovici's entropy, Parresol and Edwards' entropy and Altieri's entropy. The package is able to work with both areal and point data.

This paper is a general description of \texttt{SpatEntropy}, as well as its necessary theoretical background, and an introduction for new users. \bigskip
\end{quotation}

\begin{quotation}
\bigskip
\end{quotation}

\section[Introduction: entropy and spatial entropy in R]{Introduction: entropy and spatial entropy in \texttt{R}} \label{sec:intro}

\texttt{SpatEntropy} is the first \texttt{R} package allowing to compute entropy measures for spatial data. In applied sciences, data heterogeneity is often evaluated via computation of entropy. Entropy \citep{shan} comes from Information Theory \citep{coverthomas}, but is often employed in many statistical contexts because of its ability to synthesize different concepts such as information, surprise, uncertainty, heterogeneity, contagion; moreover, entropy indices can be constructed on any kind of variables, even unordered qualitative ones, since the computation only involves the probability of occurrence of each category. For these reasons, fields such as geography, ecology, biology and landscape studies usually refer to entropy for data description and interpretation \citep{frosini}. Often, these disciplines deal with spatial data, i.e., data that are georeferenced as points or areas. In such contexts, entropy measures should include spatial information; therefore, a number of works are available in the literature, aiming at building a spatial entropy index. They can be ascribed to three main approaches. The first starts with \cite{batty74, batty76,  batty10} who defines a spatial entropy measure which evaluates the distribution of an event over an area, allowing for unequal space partition into sub-areas. Later, \cite{karlstrom}  modified the initial proposal in order to satisfy the property of additivity in terms of decomposition of the global index into local components, following LISA criteria \citep{anselin}. The second approach to spatial entropy includes space based on a suitable transformation of the study variable to account for the distance between realizations (co-occurrences); the first proposal is made by \cite{oneill} for contiguous couples of realizations, extended by \cite{leibovici09} and \cite{leibovici14} to further distances and general degrees of co-occurrences. Contagion indices \citep{contagion,parresol} are also based on this view: spatial contagion is the opposite of entropy. As for the third approach, a set of spatial entropy measures has been presented by \cite{nostro}, starting from the co-occurrence approach but overcoming some undesirable features of the previous measures. According to this framework, Shannon's entropy of the transformed variable is decomposed into the information due to space and the remaining information brought by the transformed variable once space is considered. The proposal solves the problem of preserving additivity and disaggregating results, allowing for partial and global syntheses.

In \texttt{R} \citep{R}, the main available packages for standard entropy measures are \texttt{entropart}, \texttt{entropy} and \texttt{EntropyEstimation}. They allow basic entropy computation and decomposition into its two terms mutual information and conditional entropy \citep{coverthomas}. A quick comparison to these packages is in Section \ref{sec:shan_data}. There was no package available for spatial entropy measures, which translates the above concept into a spatial context. This work illustrates how to make use of the new package \texttt{SpatEntropy} \citep{spatentropy}, which collects functions for all the spatial entropy indices mentioned above. \texttt{SpatEntropy} makes extensive use of \texttt{spatstat} functions and data structures \citep{spatstat}. The package is built in such a way that it can be used by non-statisticians, provided they have basic knowledge of \texttt{R}; the minimum effort is requested to the user.

The following Section introduces basic notions regarding Shannon's entropy and the data examples used throughout the paper. Then, the article is organized into three Sections for the three main branches of spatial entropy measures: Section \ref{sec:batty} refers to Batty's approach; Section \ref{sec:oneill} gives details for O'Neill's approach; Section \ref{sec:nostra} illustrates Altieri's approach. Each Section can be read independently: it starts with the essential theoretical background, and then guides the reader through worked out examples covering both areal and point datasets.

\section{Data examples and entropy basics} \label{sec:shan_data}

Two datasets are used for illustration along the paper, both available within \texttt{SpatEntropy}.

\begin{figure}[t!]
\centering
\includegraphics[width=.4\textwidth]{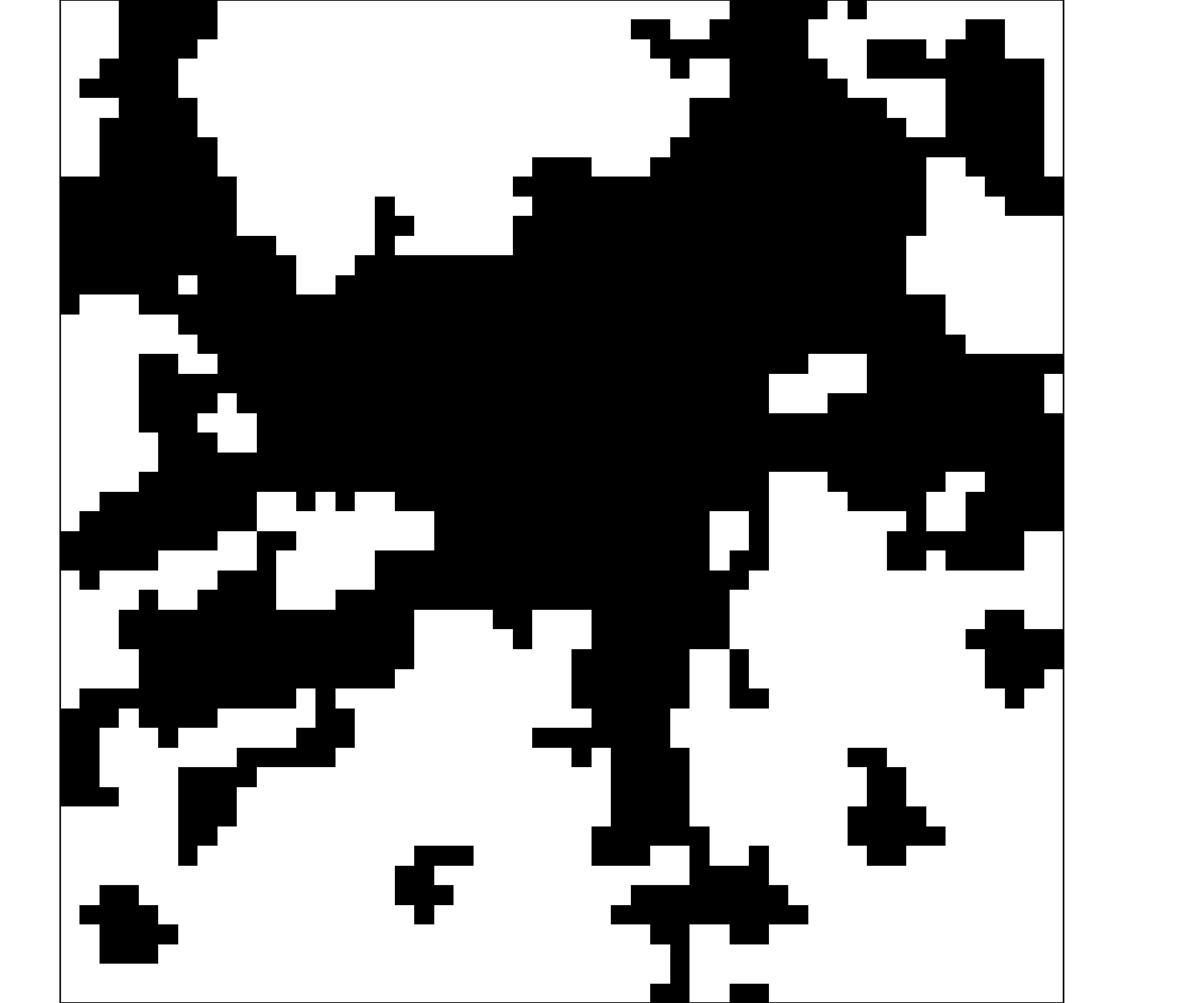}
\caption{\label{fig:bo} Trimmed Bologna urban data.}
\end{figure}

The areal dataset \texttt{data\_bologna} comes from the EU CORINE Land Cover project \citep{eea} dated 2011. It classifies the original land cover data into urbanised and non-urbanised zones, known as 'Urban Morphological Zones' (UMZ). UMZ data are useful to identify shapes and patterns of urban areas, and thus to detect what is known as urban sprawl \citep{nostro-ecolin}. Bologna's metropolitan area is extracted from the European CORINE dataset and is composed by the municipality of Bologna and the surrounding municipalities. The dataset is made of $120\times 96$ pixels of size $250\times 250$ metres and is shown in Figure~\ref{fig:bo}, where a black pixel is urban and a white pixel is non-urban. In order to speed up the results for the examples, the present paper uses a trimmed version of the original dataset: \texttt{boData}, with $n=50\times 50=2500$ cells.
\\

\noindent\texttt{R> boData=data\_bologna[41:90, 26:75]\\
R> plot\_lattice(boData, ribbon=F)}\\

The \texttt{SpatEntropy} function \texttt{plot\_lattice} allows to easily produce a gray scale map given a matrix of categorical data and, optionally, the observation area. It ensures that data is displayed following the matrix order (where position \texttt{[1,1]} corresponds to the top-left corner of the plot), avoiding risks of row inversion or transposition. A few options may be tuned, such as the extent of the gray scale, the title and the colour side legend.

\begin{figure}[t!]
\centering
\includegraphics[width=.6\textwidth]{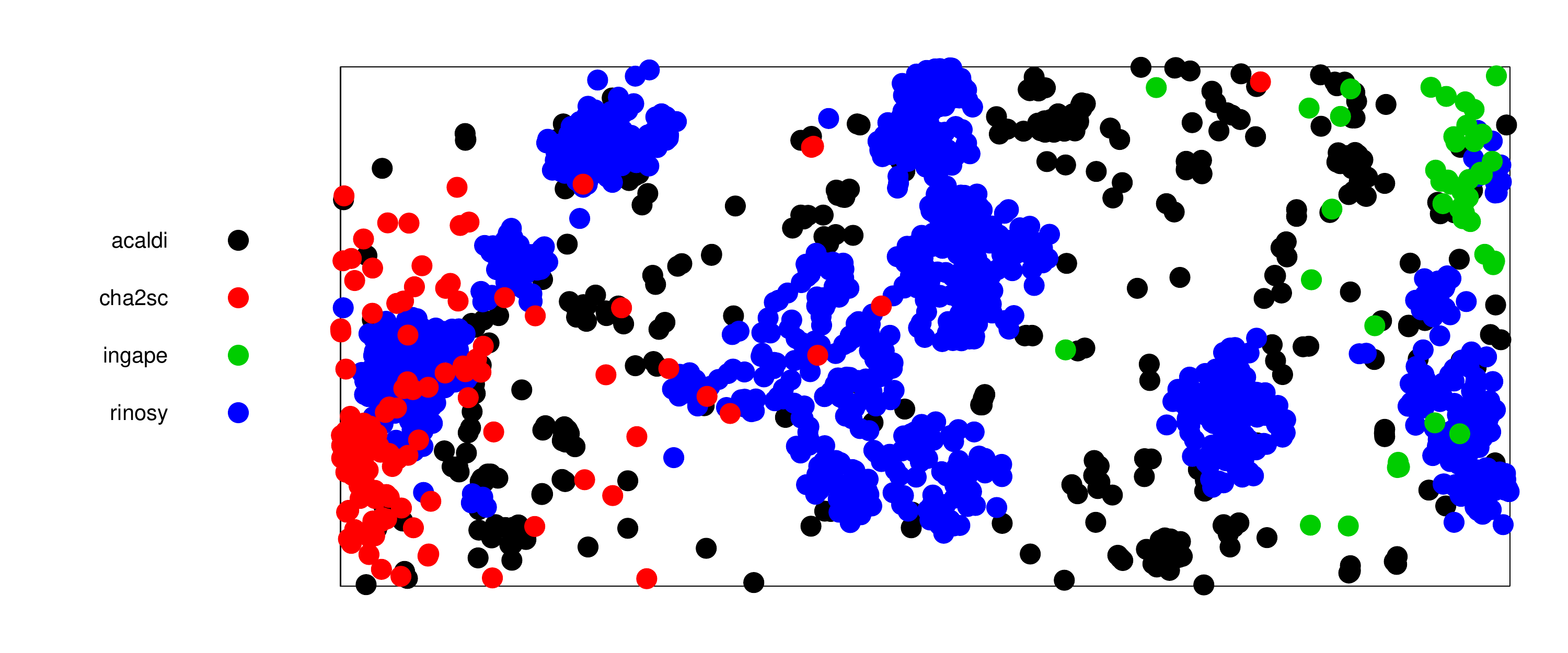}
\caption{\label{fig:rain} Trimmed rainforest tree data.}
\end{figure}

The second example dataset is \texttt{data\_rainforest}, a marked point pattern dataset about four rainforest tree species. This dataset documents the presence of tree species over Barro Colorado Island, Panama. Barro Colorado Island has been the focus of intensive research on lowland tropical rainforest since 1923 (\texttt{http://www.ctfs.si.edu}). Research identified several tree species over a rectangular observation window of size $1000\times 500$ metres; the tree species constitute the point data categorical mark. This dataset presents 4 species with different spatial configurations: Acalypha diversifolia, Chamguava schippii, Inga pezizifera and Rinorea sylvatica. The overall dataset has a total number of 7251 points. The dataset is analyzed with spatial entropy measures in \citep{nostro}. In the present article, we propose a trimmed version of the rainforest tree dataset: \texttt{treeData} with $n=1982$ trees, shown in Figure \ref{fig:rain}.\\

\noindent\texttt{R> smallW=owin(xrange=c(350,800), yrange=c(300,500))\\
R> treeData=data\_rainforest[smallW]\\
R> plot.ppp(treeData, cols=1:4, pch=19)}\\

The function \texttt{owin} belongs to \texttt{spatstat} and builds an observation area of fixed size, that we use for selecting a subset of the dataset. The function \texttt{plot.im} also comes from \texttt{spatstat} and is needed for plotting a point pattern object (an object of class \texttt{ppp}).

As a starting illustrative point, we compute Shannon's entropy for the two datasets. 
Let $X$ be a discrete random variable taking values $x_i$ in a set of $I$ outcomes. In the first above example $X$ classifies soil in $x_1$ 'urban' and $x_2$ 'non-urban' for Bologna, while for the rainforest trees $X$ is the tree species with $I=4$ categories $x_1$ to $x_4$. Let $p_X=(p(x_1),\dots,p(x_{I}))'$ be the probability mass function (pmf) of $X$. 
Shannon's entropy of $X$ is defined as 
\begin{equation}
H(X)=\sum_{i=1}^{I} p(x_i)\log\left(\frac{1}{p(x_i)}\right).
\label{eq:shann}
\end{equation}
Entropy quantifies the average amount of information brought by $X$ according to the pmf $p_X$; it is the expected value of the information function $I(p_X)$, where $I(p(x_i))=\log(1/p(x_i))$. Intuitively, outcomes with a very low probability of occurrence increase the entropy value, while outcomes very likely to occur give a small contribution to entropy. Thus, entropy measures the information coming from observing realizations, or, in other words, the surprise, which is larger when outcomes are observed that are not likely to occur. Entropy ranges in $[0,\log(I)]$ and its maximum value is achieved when $X$ is uniformly distributed. 

In \texttt{SpatEntropy}, Shannon's entropy can be computed for a dataset with \texttt{shannonX}, a function which takes matrices or vectors of any type as inputs, and returns estimated probabilities (frequencies) for all data categories together with Shannon's entropy of the dataset.\\

\noindent\texttt{R> shan.bo=shannonX(boData)}\\

\noindent\texttt{\$probabilities\\
  category frequency\\
1        0 0.5178777\\
2        1 0.4821223}\\

\noindent\texttt{\$shannon\\
{[}1{]} 0.6925078}\\

If the dataset is a \texttt{ppp} object such as \texttt{treeData}, the input of \texttt{shannonX} is the vector of point marks, i.e., the vector with the tree species.  \\

\noindent\texttt{R> shan.tree=shannonX(marks(treeData))}\\

\noindent\texttt{\$probabilities\\
  category  frequency\\
1   acaldi 0.19374369\\
2   cha2sc 0.06205853\\
3   ingape 0.02320888\\
4   rinosy 0.72098890}\\

\noindent\texttt{\$shannon\\
{[}1{]} 0.8136769}\\

In many situations, entropy is seen as a descriptive measure. It can also be seen as an estimator $\widehat{H}(X)$, where the probability distribution is estimated by the so-called plug-in estimator \citep{paninski}, which is the nonparametric as well as the maximum likelihood estimator: $\widehat{p}(x_i)=n_i/n$ substitutes the probabilities with the observed proportions over $n$ realizations. Such estimator has known properties \citep{paninski}, and its variance is 
\begin{equation}
V[\widehat{H}(X)]=V[I(p_X)]=\widehat{H}(X)^{(2)}-(\widehat{H}(X))^2,
\end{equation}
where 
$H(X)^{(2)}=\sum_{i=1}^{I} p(x_i)\log\left(\frac{1}{p(x_i)}\right)^2$.
For computing such variance, a useful function is available, \texttt{shannonX\_sq}, which computes $\widehat{H}(X)^{(2)}$:\\

\noindent\texttt{R> shan.bo2=shannonX\_sq(boData)\\
R> Vshan.bo=shan.bo2\$shannon.square-(shan.bo\$shannon\textasciicircum2)\\
R> Vshan.bo}\\

\noindent\texttt{{[}1{]} 0.001277909}\\

\noindent\texttt{R> shan.tree2=shannonX\_sq(marks(treeData))\\
R> Vshan.tree=shan.tree2\$shannon.square-(shan.tree\$shannon\textasciicircum2)\\
R> Vshan.tree}\\

\noindent\texttt{{[}1{]} 0.7451366}\\

The variance of entropy, seen as an estimator, is small when the dataset is large. 

Shannon's entropy can also be computed with the packages \texttt{entropart}, \texttt{entropy} and \texttt{Entropy\\Estimation}. Using the Bologna data example,  the function \texttt{Shannon} of \texttt{entropart} and the function \texttt{entropy.plugin} of \texttt{entropy} return the same value for $H(X)=0.6925078$, but they are less user-friendly since they require the estimates for the probabilities of all $X$ categories, while \texttt{shannonX} relies on raw data. The package \texttt{EntropyEstimation} cannot be used for comparison since it only computes the class of entropy estimators proposed by \cite{zhang}.

A major drawback of Shannon's entropy is that it does not account for the spatial location of occurrences, so that datasets with identical (estimated) pmf but very different spatial configurations share the same entropy value. The following Sections present, both in theory and practice, the three main approaches to building spatial entropy measures. Most of these measures imply the formal definition of a neighbourhood \citep{cressie}. The simplest way of representing a neighbourhood system over $n$ spatial units is via an adjacency matrix \citep{anselin}, i.e., a square matrix whose elements indicate whether pairs of units are neighbours: $a_{uu'}=1$ if $u' \in \mathcal{N}(u)$, that is the neighbourhood of area $u$; $a_{uu}=0$ by definition. Spatial units may be points, defined via coordinate pairs, or areas, identified via representative coordinate pairs, such as the area centroids.

\section{Spatial entropy for an area partition} \label{sec:batty}

Batty's spatial entropy \citep{batty74, batty76, batty10} is useful for evaluating the heterogeneity in the distribution of a phenomenon over an area. It is particularly appropriate when the observation area is exogenously partitioned into sub-areas (such as municipality administrative boundaries for a Region). If the probabilities of the neighbouring sub-areas should enter the computation, one should resort to the proposal by \cite{karlstrom}.

\subsection{Batty's entropy}

Let a phenomenon of interest $F$ occur over an observation window of size $T$ partitioned into $G$ areas of size $T_g$. This defines $G$ dummy variables identifying the occurrence of $F$ over a generic area $g$, $g=1, \dots, G$.
Given that $F$ occurs over the window, its occurrence in area $g$ takes place with probability $p_g$, where $1-p_g=\sum_{g' \ne g} p_{g'}$ and $\sum_g p_g=1$. The phenomenon intensity is obtained as $\lambda_g=p_g/T_g$, where $T_g$ is the area size, and is assumed constant within each area. Batty's spatial entropy is
\begin{equation}
H_B(F)=\sum_{g=1}^G p_g \log \left(\frac{T_g}{p_g}\right).
\label{eq:spaten}
\end{equation}
It expresses the average amount of information brought by the occurrence of $F$ in any area in the observation window, and includes a multiplicative component $T_g$ that accounts for unequal space partition. Analogously to Shannon's entropy, which is high when the $I$ categories of $X$ are equally represented over a (non spatial) data collection, Batty's entropy is high when the phenomenon of interest $F$ is equally intense over the $G$ areas partitioning the observation window (i.e., when $\lambda_g=\lambda$ for all $g$). Batty's entropy $H_B(F)$ reaches a minimum value equal to $\log(T_{g^*})$ when $p_{g^*}=1$ and $p_g=0$ for all $g\ne g^*$, with $g^*$ denoting the area with the smallest size. The maximum value of Batty's entropy is $\log(T)$, reached when the intensity of $F$ is the same over all areas, i.e., $\lambda_g=1/T$ for all $g$.

\subsection[Batty's entropy with SpatEntropy]{Batty's entropy with \texttt{SpatEntropy}}
\label{sec:battyR}

The key function for computing Batty's entropy in \texttt{SpatEntropy} is\\

\noindent\texttt{batty(data, data.assign, is.pointdata = FALSE, category, win = NULL, G.coords)}\\

\noindent where \texttt{data} can be a data matrix or vector of any type. The two arguments \texttt{data.assign} and \texttt{G.coords} summarize the information concerning the partition into sub-areas: \texttt{data.assign} matches each spatial unit to the sub-area with the closest centroid, while  \texttt{G.coords} contains the coordinates of the sub-areas centroids. They can be obtained as the output of another function of \texttt{SpatEntropy}, \texttt{areapart}:\\

\noindent\texttt{areapart(win, G, data.coords)}\\

The function \texttt{areapart} needs \texttt{win}, the observation area as an \texttt{owin} object (see \texttt{spatstat} and the examples in the following Sections), \texttt{G} ruling the partition into sub-areas, and \texttt{data.coords} as a two-column matrix. The argument \texttt{G} is a two-column matrix with the sub-areas' centroid coordinates if a fixed area partition is provided, or an integer determining the number of sub-areas if they are randomly generated within the function. We recommend to provide an meaningful exogenous area partition, since conclusions for Batty's entropy are heavily affected by the partition itself. The output of \texttt{areapart} is a three-column matrix named \texttt{data.assign} which associates an area id to each spatial unit coordinate pair, and \texttt{G.coords}, the partition in sub-areas. This is part of the input of \texttt{batty}. Other arguments of \texttt{batty} are discussed separately for areal and point data in the following.

The output of \texttt{batty} is the value for Batty's entropy, a single number, and a table summarizing information about the phenomenon under study. Information is provided about each sub-area: \texttt{area.id}, the sub-area id, \texttt{abs.freq}, the number of points/pixels presenting the category of interest for each sub-area, \texttt{rel.freq}, the relative frequency which is used as an estimate for the probability $p_g$, and \texttt{Tg}, the sub-area size.

\subsubsection{Areal data}

The workflow for computing Batty's entropy from scratch for areal data, taking Bologna data as an example, is
\begin{itemize}
\item create the data observation window. Without loss of generality, we can assume Bologna's pixels are of size 1 and thus create a square of size $50 \times 50$\\

\noindent\texttt{R> bo.win=owin(xrange=c(0, ncol(boData)),\\ yrange=c(0,nrow(boData)))}

\item find the units' coordinates. The \texttt{SpatEntropy} function \texttt{coords\_pix} is thought for lattice data and provides the centroids coordinates for all pixels
\\

\noindent\texttt{R> bo.cc=coords\_pix(bo.win, pixel.xsize=1, pixel.ysize=1) }\\

As an alternative to the two dimensions of the pixel \texttt{pixel.xsize} and \texttt{pixel.ysize}, the number of rows and columns of the grid can be given as \texttt{nrow} and \texttt{ncol}. Note that \texttt{nrow} is the number of pixels along the $y$ axis of the plot, and \texttt{ncol} is the number of pixels along the $x$ axis
\item partition the observation window into $G$ sub-areas \\

\noindent\texttt{R> bo.part=areapart(bo.win, G=4, data.coords=bo.cc)}\\

In this case, $G=4$ means four random centroids are generated over the window. Then, the sub-areas borders are built following the pixel borders and assign each pixel to the closest area centroid. The area partition can be plotted as in Figure \ref{fig:bo_part} with\\

\noindent\texttt{R> plot\_areapart(bo.part\$data.assign, bo.win, is.pointdata=F, add.data=T,\\
+  data=boData, G.coords=NULL, main="")}\\

The input \texttt{G.coords} is not needed for lattice data; the option \texttt{is.pointdata} is set to \texttt{FALSE} for areal data, while the option \texttt{add.data} indicates whether to plot only the area partition (\texttt{add.data=F}, Figure~\ref{fig:bo_part}, left panel), or to plot it together with the data (\texttt{add.data=T}, Figure~\ref{fig:bo_part}, right panel)
\begin{figure}[t!]
\centering
\includegraphics[width=.6\textwidth]{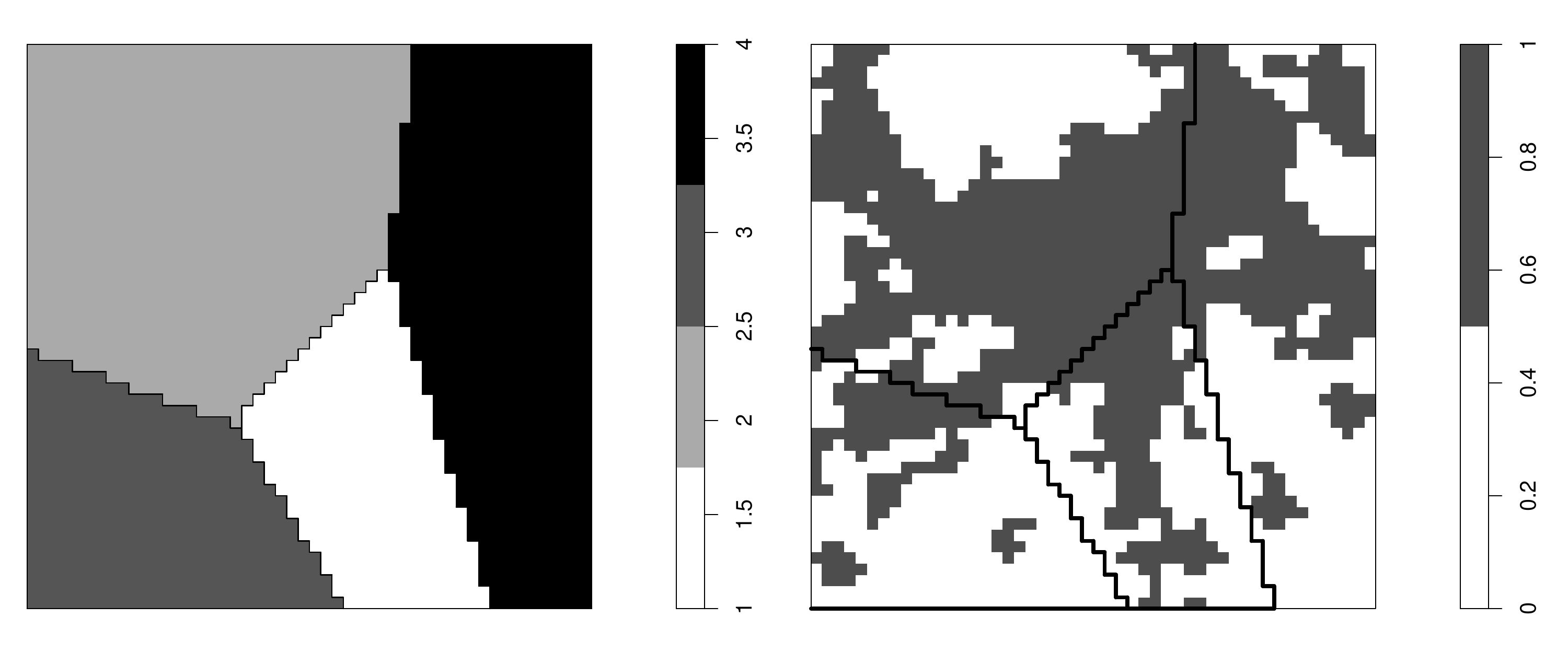}
\caption{\label{fig:bo_part} Partition into 4 random sub-areas for Bologna dataset.}
\end{figure}
\item compute Batty's entropy for the phenomenon of interest, which may be "urban pixels" or "non-urban pixels"; one value for Batty's entropy may be computed for each category of the variable, by specifying the argument \texttt{category}.\\

\noindent\texttt{R> bo.batty1=batty(boData, bo.part\$data.assign, category=1, \\
+  win=bo.win, G.coords=bo.part\$G.coords)\\
R> bo.batty1\$batty.ent}\\

\noindent\texttt{{[}1{]} 7.788642}\\

\noindent\texttt{R> bo.batty0=batty(boData, bo.part\$data.assign, category=0, \\
+  win=bo.win, G.coords=bo.part\$G.coords)\\
R> bo.batty0\$batty.ent}\\

\noindent\texttt{{[}1{]} 7.795131}
\end{itemize}

\subsubsection{Point data}

In this Section, the differences in the functions are highlighted between areal and point data, and between binary and multicategorical data. The workflow for computing Batty's entropy for point pattern data with categorical marks, taking rainforest tree data as an example, is
\begin{itemize}
\item find the points' coordinates, by exploiting \texttt{coords.ppp} from \texttt{spatstat} \\

\noindent\texttt{R>tree.cc=coords.ppp(treeData) }\\

\item partition the observation window into $G$ sub-areas \\

\noindent\texttt{R> tree.part=areapart(treeData\$win, G=6, data.coords=tree.cc)}\\

\noindent where $G=6$ means six random centroids are generated over the window. The sub-areas borders are built following the Dirichlet tessellation (see \texttt{?dirichlet}, a \texttt{spatstat} function), i.e. points are assigned to the area with the closest centroid. The area partition can be plotted with\\

\noindent\texttt{R> plot\_areapart(bo.part\$data.assign, bo.win, is.pointdata=T, add.data=T,\\
+  data.bin=T, category="rinosy", data=treeData,\\
+  G.coords=tree.part\$G.coords, main="")}

\noindent where the option \texttt{is.pointdata} is set to \texttt{TRUE}. If the option \texttt{add.data} for displaying points is set to \texttt{TRUE}, for multicategorical data one can choose whether to plot all points or to select a category of interest. In the former case, \texttt{data.bin=F} (Figure~\ref{fig:tree_part}, left panel), while in the latter case \texttt{data.bin=T} (Figure~\ref{fig:tree_part}, right panel) for data dichotomization according to the specified \texttt{category}
\begin{figure}[t!]
\centering
\includegraphics[width=\textwidth]{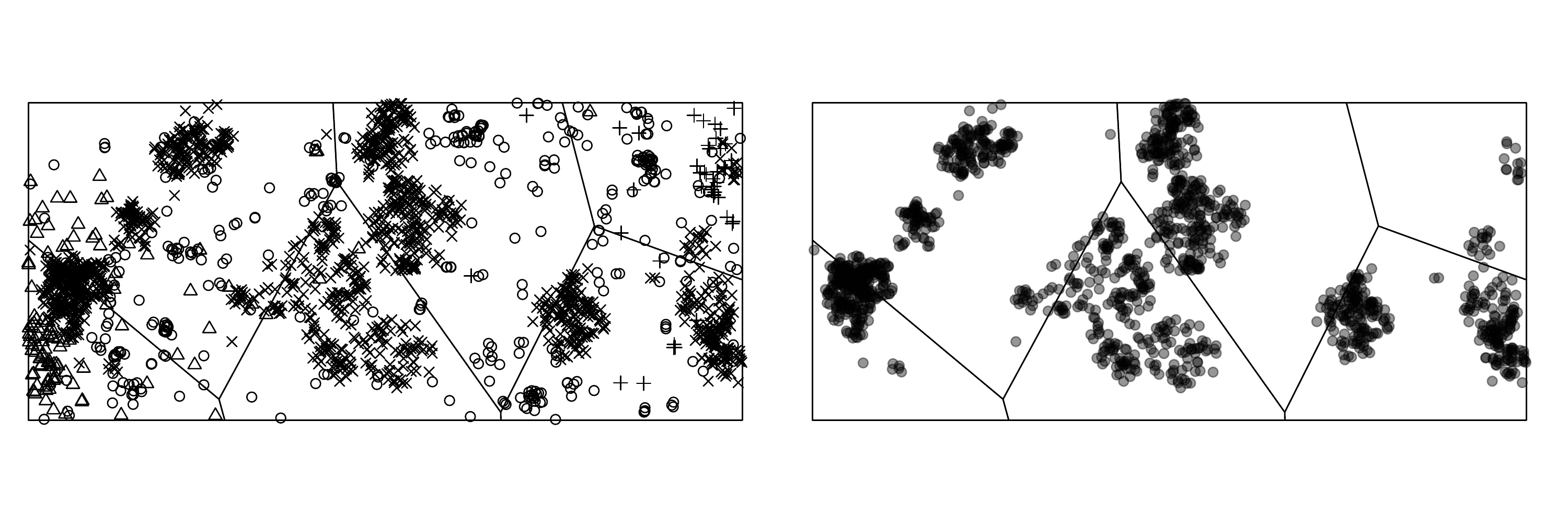}
\caption{\label{fig:tree_part} Partition into 6 random sub-areas for rainforest tree dataset. Left panel: all trees are plotted; right panel: only the specie Rinorea sylvatica is plotted.}
\end{figure}
\item compute Batty's entropy for a category of interest, with the option \texttt{is.pointdata=T}\\

\noindent\texttt{R> tree.batty.rinosy=batty(marks(treeData), tree.part\$data.assign,\\
+  is.pointdata=T, category="rinosy", win=treeData\$win, \\
+  G.coords=tree.part\$G.coords)\\
R> tree.batty.rinosy\$batty.ent}\\

\noindent\texttt{{[}1{]} 11.32699}\\
\end{itemize}

\subsection{Karlstr\"om and Ceccato's entropy}

A challenging attempt to introduce additive properties and to include the idea of neighbourhood in Batty's entropy index (\ref{eq:spaten}) is due to \cite{karlstrom}, following the LISA theory \citep{anselin}. 
The adjacency matrix for $G$ spatial units is $A=\{a_{gg'}\}_{g,g'=1,\dots,G}$, a square $G\times G$ matrix such that $a_{gg'}=1$ when $g' \in \mathcal{N}(g)$, the neighbourhood of area $g$. In this proposal, the elements on the diagonal of the adjacency matrix $A$ are non-zero, i.e., each area neighbours itself. 

\citeauthor{karlstrom}'s entropy index $H_{KC}(F)$ starts by weighting the probability of occurrence of $F$ in a given spatial unit $g$, $p_g$, with its neighbouring values:
\begin{equation}
\widetilde{p}_g=\sum_{g'=1}^G a_{gg'}p_{g'}.
\label{eq:karl_ptilde}
\end{equation}
Then, an information function is defined, fixing $T_g=1$, as $I(\widetilde{p}_g)=\log\left(1/\widetilde{p}_g\right)$. \citeauthor{karlstrom}'s entropy index is
\begin{equation}
H_{KC}(F)=E\left[I\left(\widetilde{p}_g\right)\right]=\sum_{g=1}^G p_g\log\left(\frac{1}{\widetilde{p}_g}\right).
\label{eq:karl}
\end{equation}
The maximum of $H_{KC}(F)$ does not depend on the choice of the neighbourhood and is $\log(G)$. As the neighbourhood reduces, i.e., as $A$ tends to the identity matrix, $H_{KC}(F)$ tends to Batty's spatial entropy (\ref{eq:spaten}), with equality in the case of $T_g=1$ for all $g$. The sum of local measures $L_g=p_gI(\widetilde{p}_g)$ forms the global index (\ref{eq:karl}), preserving the LISA property of additivity.

\subsection[Karlstr\"om and Ceccato's entropy with SpatEntropy]{Karlstr\"om and Ceccato's entropy with \texttt{SpatEntropy}}

This is a modified version of Batty's entropy, so one may refer to Section \ref{sec:battyR} for the necessary preamble. The key function is\\

\noindent\texttt{karlstrom(data, data.assign, category, G.coords, neigh.dist)}\\

\noindent where \texttt{batty}'s inputs \texttt{is.pointdata} and \texttt{win} are missing since they are needed for the computation of the areas $T_g$, which are discarded in this entropy measure. The only new input with regard to Batty's entropy is \texttt{neigh.dist}; this is a scalar fixed by the user, expressing the extent of the neighbourhood in all directions. The value must be chosen keeping the area size into account.

\subsubsection{Areal data}

After all the steps needed for Batty's entropy outlined in Section \ref{sec:battyR}, for Bologna data, we write\\

\noindent\texttt{R> bo.karlstr=karlstrom(boData, bo.part\$data.assign, category=1,\\
+  bo.part\$G.coords, neigh.dist=15)}\\

\noindent\texttt{\$karlstrom.table\\
     area.id abs.freq   rel.freq    p.tilde\\
{[}1,{]}       1      173 0.14076485 0.30634662\\
{[}2,{]}       2      580 0.47192840 0.30634662\\
{[}3,{]}       3      115 0.09357201 0.09357201\\
{[}4,{]}       4      361 0.29373474 0.29373474}\\

\noindent\texttt{\$karlstrom.ent\\
{[}1{]} 1.306362}\\

Setting \texttt{neigh.dist=15} means that, when estimating $\widetilde{p}_g$ of (\ref{eq:karl_ptilde}) for a sub-area $g$, the proportions in all sub-areas whose centroid is at most 15 spatial units apart enter the computation. The output is analogous to the output of \texttt{batty}.

\subsubsection{Point data}

For point data with multicategorical marks, starting from the steps of Section \ref{sec:battyR}, we write\\

\noindent\texttt{R> tree.karlstr=karlstrom(marks(treeData), tree.part\$data.assign, \\
+  category="rinosy", tree.part\$G.coords, neigh.dist=100)}\\

\noindent\texttt{\$karlstrom.table\\
     area.id abs.freq  rel.freq   p.tilde\\
{[}1,{]}       1       30 0.0209937 0.1074178\\
{[}2,{]}       2      277 0.1938418 0.1074178\\
{[}3,{]}       3      258 0.1805458 0.1805458\\
{[}4,{]}       4      201 0.1406578 0.1406578\\
{[}5,{]}       5      341 0.2386284 0.2386284\\
{[}6,{]}       6      322 0.2253324 0.2253324}\\

\noindent\texttt{\$karlstrom.ent\\
{[}1{]} 1.741951}\\

\noindent where again \texttt{neigh.dist} is chosen by the user considering the total area size.

\section{Entropy for spatially associated categorical variables} \label{sec:oneill}

A second way to build a spatial entropy measure relies on defining a new categorical variable $Z$, where each realization identifies ordered couples $(x_i,x_j)$ of occurrences of $X$ over space. Order preservation within couples regards considering the relative spatial location of the observations. If order is preserved the couple $(x_i,x_j)$ implies that the observation carrying the $j$-th category occurs at the right or below the observation carrying the $i$-th category. Under this criterion, such couple is different from $(x_j,x_i)$. For $I$ categories of $X$, the new variable $Z$ has $R=I^2$ categories. The attention moves from the computation of (\ref{eq:shann}), namely $H(X)$, to an index of the same form, Shannon's entropy of $Z$, $H(Z)$. 

The entropy measures based on $Z$ are useful when the variable of interest has two or more categories and when the goal is to understand how an outcome at one location affects neighbouring outcomes. Intuitively, when the variable is strongly spatially associated, neighbouring outcomes are closely related, which decreases the surprise (and thus, the entropy) in observing data. Such measures are based on selecting couples occurring at one specific distance; in the standard case, contiguous couples are considered, but extensions to farther distances are allowed. O'Neill and Leibovici's entropies \citep{oneill, leibovici09} quantify the residual amount of entropy associated to the variable of interest, once the influence of the spatial configuration has been taken into account at a specific distance. The chosen distance defines a neighbourhood, which is fixed prior to the analysis, and excludes information at farther distances. When the interest lies in what happens at contiguous locations, i.e., by considering areal units sharing a border, O'Neill's entropy should be computed, or one of its contagion versions. When point data are available, or when distances other than contiguity are under study, Leibovici's entropy should be used, which is a generalization of O'Neill's entropy.

\subsection{O'Neill's entropy and contagion indices}

\cite{oneill} propose one of the early spatial entropy indices for lattice data. It is based on computing a Shannon's entropy (\ref{eq:shann}) for the subset of the variable $Z$ made of contiguous couples, i.e., spatial realizations sharing a border. Such couples are identified by non-zero elements in the special adjacency matrix named contiguity matrix $C$. The subset of couples of contiguous realizations is $Z|C$, and its Shannon's entropy is
\begin{equation}
H(Z|C)=\sum_{r=1}^{I^2} p(z_{r}|C)\log\left(\frac{1}{p(z_{r}|C)}\right).
\label{eq:oneill}
\end{equation}
Entropy (\ref{eq:oneill}) ranges from 0 to $\log(I^2)$.

Other measures based on the construction of $Z|C$ start from the concept of contagion, the conceptual opposite of entropy. The Relative Contagion index $RC$ \citep{contagion} is proposed as
\begin{equation}
RC(Z|C)=1-H_{norm}(Z|C)=1-\frac{1}{\log(I^2)}\sum_{r=1}^{I^2} p(z_{r}|C)\log\left(\frac{1}{p(z_{r}|C)}\right).
\label{eq:rc}
\end{equation}
The second term is the normalized entropy of $Z|C$, via the multiplication of (\ref{eq:oneill}) by $1/\log(I^2)$. Its complement to 1 is computed in order to measure relative contagion: the higher the spatial contagion between categories of $Z|C$, the lower the spatial entropy.

If one wants to account for the number of categories of $X$ when computing the contagion index, non-normalized measures should be computed in order to distinguish among contexts with different numbers of categories. For this reason, \cite{parresol} suggest an unnormalized version of (\ref{eq:rc}):
\begin{equation}
P(Z|C)=-H(Z|C)=\sum_{r=1}^{I^2} p(z_{r}|C)\log(p(z_{r}|C))
\label{eq:gamma}
\end{equation}
thus ranging from $-\log(I^2)$ to $0$.

\subsection[O'Neill's entropy and contagion indices with SpatEntropy]{O'Neill's entropy and contagion indices with \texttt{SpatEntropy}}
\label{sec:oneillR}

The key function for O'Neill's entropy is\\

\noindent \texttt{leibovici(data, adj.mat, missing.cat = NULL, ordered = TRUE)}\\

\noindent since, as explained in Section \ref{sec:leib}, O'Neill's entropy is actually a special case of Leibovici's entropy. As usual, \texttt{data} is a data matrix or vector, which can be numeric, factor, character. The input \texttt{adj.mat} is the contiguity matrix for O'Neill's entropy and the contagion indices, i.e., a matrix identifying areal units sharing a border. The option \texttt{missing.cat} accounts for categories of the variable of interest that are absent in the dataset, while \texttt{ordered} is set as \texttt{TRUE} according to the authors' choice to consider ordered couples of realizations.

\subsubsection{Areal data and contiguity matrix}

In order to compute O'Neill's entropy and the contagion indices for Bologna lattice data, the starting point is the object \texttt{bo.cc} created in Section \ref{sec:battyR}. Then, the workflow is
\begin{itemize}
\item compute the matrix of all Euclidean distances between pixel centroids\\

\noindent\texttt{R> bo.dmat=euclid\_dist(bo.cc)}\\

Note that this \texttt{SpatEntropy} function may require some time for large datasets, but is only computed once and can then be used for any adjacency matrix and any entropy index. An alternative option is possible, which is faster though a little more complicated to program\\

\noindent\texttt{R> bo.ccP=ppp(bo.cc{[},1{]}, bo.cc{[},2{]}, bo.win)\\
R> bo.dmat=pairdist(bo.ccP)\\
R> bo.dmat[lower.tri(bo.dmat,diag=TRUE)] <- NA}\\

\noindent This option exploits \texttt{spatstat} functions: it turns the set of coordinates into a \texttt{ppp} object and uses the function \texttt{pairdist} to compute the distance between all pairs of centroids. Then, it turns the resulting symmetric matrix into a more efficient upper-triangular matrix
\item build the contiguity matrix\\

\noindent\texttt{R> bo.adjmat=adj\_mat(bo.dmat, dd0=0, dd1=1)}\\

\noindent where \texttt{dd0} is the minimum distance and is always set to 0 for O'Neill's entropy, while \texttt{dd1} is equal to the pixel width in order to select only couples of pixels sharing a border (i.e., contiguous). The \texttt{SpatEntropy} function \texttt{adj\_mat} builds an upper-triangular adjacency matrix, which means that couples of pixels are counted moving downward and rightward along the observation window. This ensures computational efficiency and avoids double counting of couples
\item compute O'Neill's entropy\\

\noindent\texttt{R> bo.oneill=leibovici(boData, bo.adjmat, ordered=T)}\\

This function makes use of the \texttt{SpatEntropy} function \texttt{couple\_count} when \texttt{ordered=T}, (or the analogous \texttt{pair\_count} when \texttt{ordered=F}) for building all possible adjacent couples in the dataset and computing the relative frequencies, which enter the computation of O'Neill's entropy as estimates of the probabilities. The function \texttt{leibovici} returns a summary of the data structure, and the value of O'Neill's entropy\\

\noindent\texttt{\$freq.table\\
  couple abs.frequency proportion\\
1     00          2159 0.44061224\\
2     01           298 0.06081633\\
3     10           315 0.06428571\\
4     11          2128 0.43428571}\\

\noindent\texttt{\$entropy\\
{[}1{]} 1.070045}\\
\end{itemize}
The Relative Contagion index and Parresol and Edwards' contagion can be computed in a similar way with the functions\\

\noindent\texttt{contagion(oneill = NULL, n.cat = NULL, data = NULL, adj.mat = NULL,\\ 
          \indent missing.cat = NULL, ordered = TRUE)\\
parresol(oneill = NULL, n.cat = NULL, data = NULL, adj.mat = NULL, \\
        \indent missing.cat = NULL, ordered = TRUE)}\\
         
The starting point for these indices may alternatively be \texttt{oneill}, the special output of \texttt{leibovici} as above, or the single raw inputs the same way as for \texttt{leibovici}. The only new input is \texttt{n.cat} which is the number of categories of the variable under study.

\subsubsection{Point data and other computational options}

The function is very flexible and allows to tune some options for computing entropy in an extended way. It can be used for point data; in such case, \texttt{data} is the mark vector. It allows for adjacency matrices different than the contiguity matrix, and for unordered couples (i.e., pairs). Some of these options are explored in Section \ref{sec:leib}. The possibility to work with pairs instead of couples is discussed in Section \ref{sec:nostra}; for interpretation, remember that, if pairs are chosen instead of couples, the entropy value is smaller since the number of possible categories for $Z$ is smaller.

\subsection{Leibovici's entropy}
\label{sec:leib}

\cite{leibovici09} and \cite{leibovici14} propose a richer measure of entropy by extending $H(Z|C)$ in two ways. Firstly, $Z$ can now represent not only couples, but also triples and further degrees $m$ of co-occurrences. The authors develop the case of ordered co-occurrences, so that the number of categories of $Z$ is $R_m=I^m$. Secondly, space is now allowed to be continuous, so that areal as well as point data might be considered and associations may not coincide with contiguity: the concept of distance between occurrences replaces the concept of contiguity between lattice cells. A distance $d$ is fixed, then co-occurrences are defined for each $m$ and $d$ as $m$-th degree simultaneous realizations of $X$ at any distance $d^{*} \le d$, i.e., distances are considered according to a cumulative perspective; this way an adjacency hypercube $A_d$ is built and the subset of interest is $Z|d$. Then, Leibovici's spatial entropy is
\begin{equation}
H(Z|d)=\sum_{r=1}^{I^m} p(z_r|d)\log\left(\frac{1}{p(z_r|d)}\right).
\label{eq:leib}
\end{equation}
In the case of lattice data, O'Neill's entropy (\ref{eq:oneill}) is obtained as a special case when $m=2$ and $d$ equals the cell's width.

\subsection[Leibovici's entropy with SpatEntropy]{Leibovici's entropy with \texttt{SpatEntropy}}

Leibovici's entropy is currently implemented for couples ($m=2$) with the function \texttt{leibovici}. The key difference with regard to the examples of Section \ref{sec:oneillR} is the possibility to choose the adjacency distance and to work with point data, not only with areal data.

\subsubsection{Areal data}

Leibovici's entropy for Bologna data can be computed in a similar way to what illustrated in Section \ref{sec:oneillR} for a generic distance\\

\noindent\texttt{R> bo.adjmat5=adj\_mat(bo.dmat, dd0=0, dd1=5)\\
R> bo.leib=leibovici(boData, bo.adjmat5, ordered=T)}\\

\noindent where the value for the distance range of interest \texttt{dd1} is set by the user. The output is analogous to the case in Section \ref{sec:oneillR}.

\subsubsection{Point data}

\begin{figure}[t!]
\centering
\includegraphics[width=.5\textwidth]{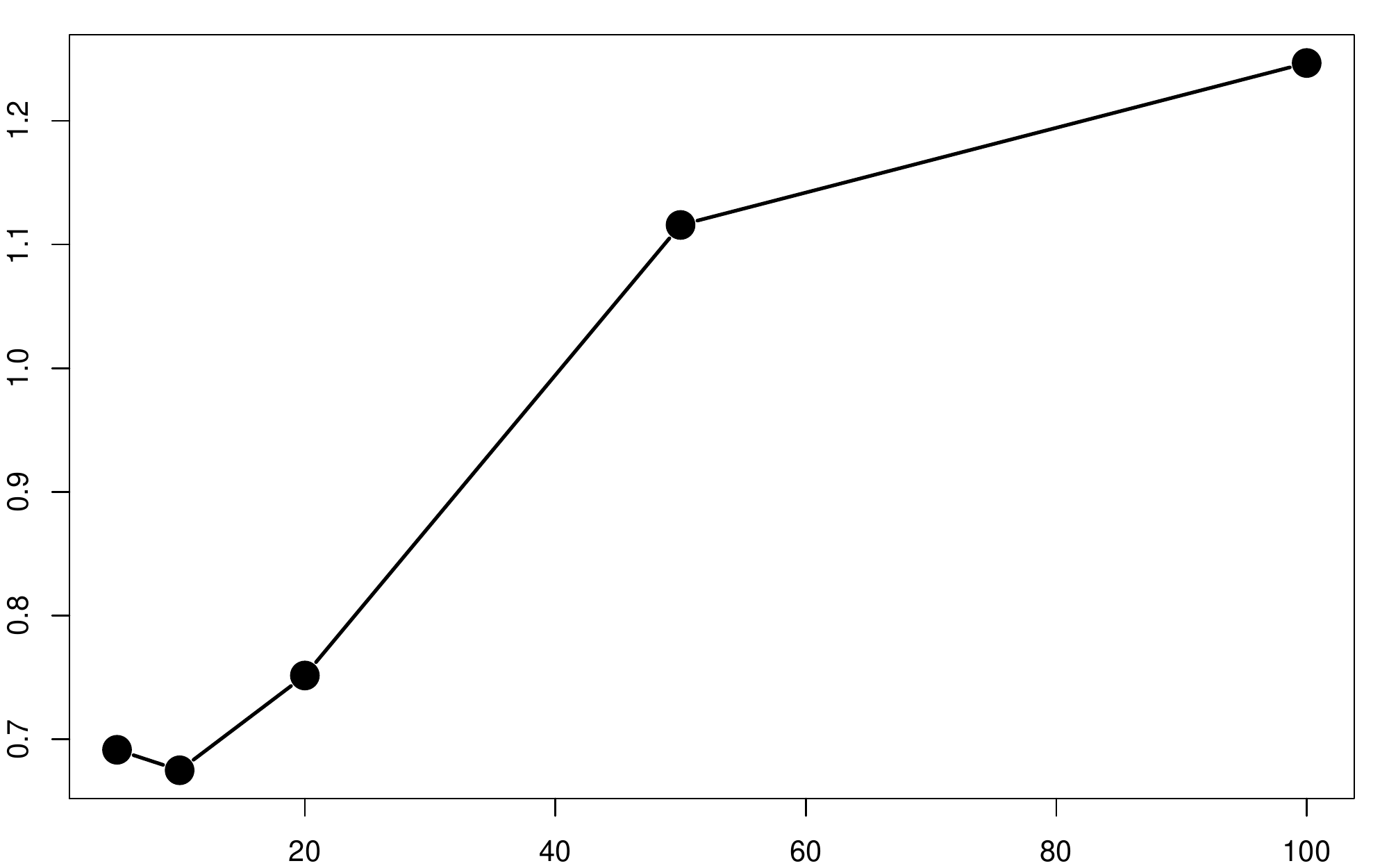}
\caption{\label{fig:leibs} Leibovici's entropy for tree data with $d=(5, 10, 20, 50, 100)$.}
\end{figure}

Leibovici's entropy works the same way as for areal data for point data: a maximum distance range is chosen for the adjacency matrix, then couples are built by looking for points that lie within the distance range. The workflow, with the rainforest tree data example, is
\begin{itemize}
\item compute the matrix of all Euclidean distances between points (exploiting a \texttt{spatstat} function)\\

\noindent\texttt{R> tree.dmat=pairdist(treeData)\\
R> tree.dmat[lower.tri(tree.dmat,diag=TRUE)] <- NA}\\

\item build the adjacency matrix for a chosen distance\\

\noindent\texttt{R> tree.adjmat=adj\_mat(tree.dmat, dd0=0, dd1=20)}\\

\noindent where \texttt{tree.adjmat} is an upper-triangular adjacency matrix identifying all couples of trees at distance \texttt{dd1} or less apart; \texttt{dd1} is set by the user
\item compute Leibovici's entropy\\

\noindent\texttt{R> tree.leib=leibovici(marks(treeData), tree.adjmat, ordered=T)\\
R> tree.leib\$entropy}\\

\noindent\texttt{{[}1{]} 0.7516415}\\

\end{itemize}
Several Leibovici's entropy values for different choices of \texttt{dd1} may be compared; an example for the tree data is shown in Figure~\ref{fig:leibs}.

\section{A decomposable spatial entropy measure} \label{sec:nostra}

This is the most recent approach for spatial entropy measures. It should be employed when the interest lies in understanding the role of the spatial configuration in determining the entropy of a variable, not only at one isolated specific distance but also at a global level or at different distance ranges simultaneously. In addition, it should be used when the influence of space needs to be quantified as a percentage of the entropy. It is a more sophisticated approach from a statistical perspective, and allows more flexibility and interpretability than the previous measures. 

The starting point is a different way of computing the variable $Z$ of Section \ref{sec:oneill}; it is extendable to a general degree $m$ of co-occurrences (see Section \ref{sec:leib}), but it is currently implemented for pairs of realizations in \texttt{SpatEntropy}.

\subsection{Altieri's entropy}

\cite{nostro} follow the approach based on $Z$  discarding order within co-occurrences, meaning that the relative spatial location of the two realizations is irrelevant; therefore, pairs are considered instead of couples. Discarding the order ensures a one-to-one correspondence between Shannon's entropy of $X$ and $Z$. Moreover, ordering occurrences is not sensible in spatial statistics, where spatial configurations are not generally assumed to have a direction. Besides, when order is discarded, the number of categories of $Z$ is smaller. The gap between the two options grows as $I$ increases, and induces a different computational burden for large datasets. 

A second discrete variable $W$ is introduced, that represents space by classifying the distances at which the two occurrences of a pair take place. Classes $w_k$ must be defined, with $k=1,\dots,K$, covering all possible distances within the observation window. Each distance class $w_k$ implies the choice of a corresponding adjacency matrix $A_k$, which identifies pairs where the two realizations of $X$ lie at a distance belonging to the range $w_k$. 

Thanks to the introduction of $W$, the entropy of $Z$ 
\begin{equation}
H(Z)=\sum_{r=1}^R p(z_r)\log\left(\frac{1}{p(z_r)}\right)=MI(Z,W)+H(Z)_{W}
\label{eq:shannZ_add}
\end{equation}
may be decomposed following the basis of Information Theory \citep{coverthomas}. In (\ref{eq:shannZ_add}), the two terms acquire a spatial meaning: $MI(Z,W)$ is spatial mutual information, quantifying the part of entropy of $Z$ due to the spatial configuration $W$; $H(Z)_{W}$ is spatial global residual entropy, quantifying the remaining information brought by $Z$ after space has been taken into account. The more $Z$ depends on $W$, i.e. the more the realizations of $X$ are spatially associated, the higher the spatial mutual information. Conversely, when the spatial association among the realizations of $X$ is weak, the entropy of $Z$ is mainly due to spatial global residual entropy. The entropy $H(Z)$ is a stable reference value, while its two components $MI(Z,W)$ and $H(Z)_W$ vary in order to evaluate the role of space for datasets with different spatial configurations. This is only the case when order is discarded. For the sake of interpretation and diffusion of the results, $MI_{prop}(Z,W)=MI(Z,W)/H(Z)$ may be used, which ranges in $[0,1]$ and is able to quantify the proportional contribution of space in the entropy of $Z$. 

The overall value of $MI(Z,W)$, however, is often negatively influenced by what happens at large distance ranges, where usually scarce correlation is present. Hence, spatial mutual information for the whole dataset may be low even when a clustered pattern occurs. The variable $W$ helps in overcoming this drawback, since the two terms forming $H(Z)$ can be further decomposed. Indeed, $K$ subsets of realizations of $Z$ are available,
denoted by $Z|w_k$. Spatial mutual information
\begin{equation}
MI(Z,W)=\sum_{k=1}^{K}p(w_k)PI(Z|w_k)
\label{eq:PI_add}
\end{equation}
is a weighted sum of partial terms, where
\begin{equation}
PI(Z|w_k)=\sum_{r=1}^{R}p(z_r|w_k)\log{\left(\frac{p(z_r|w_k)}{p(z_r)}\right)}.
\label{eq:partialterm_mut}
\end{equation}
Each partial term $PI(Z|w_k)$ quantifies the contribution to the departure from independence of each conditional distribution $p_{Z|w_k}$, i.e. the contribution of the $k$-th distance range to the global mutual information between $Z$ and $W$. 
Analogously, the following additive decomposition holds:
\begin{equation}
H(Z)_W=\sum_{k=1}^{K}p(w_k)H(Z|w_k),
\label{eq:residZWadd}
\end{equation}
where the partial residual entropy terms measure the partial contributions to the entropy of $Z$ due to sources other than the spatial configuration:
\begin{equation}
H(Z|w_k)=\sum_{r=1}^{R}p(z_r|w_k)\log{\left(\frac{1}{p(z_r|w_k)}\right)} .
\label{eq:residZW_loc}
\end{equation}

\subsection[Altieri's entropy with SpatEntropy]{Altieri's entropy with \texttt{SpatEntropy}}

The function for computing Altieri's spatial entropy in \texttt{SpatEntropy} is\\

\noindent\texttt{spat\_entropy(data, adj.list, shannZ, missing.cat = NULL)}\\

\noindent which relies on two other functions:\\

\noindent\texttt{adj\_list(dist.mat, dist.breaks)}\\

\noindent and\\

\noindent\texttt{shannonZ(data, missing.cat = NULL)}\\

The function \texttt{adj\_list} builds a list of adjacency matrices for a fixed partiton into distance classes. The function \texttt{shannonZ} starts from \texttt{data}, a data matrix or vector, and computes $H(Z)$ for unordered couples of realizations, exploiting the auxiliary \texttt{SpatEntropy} function \texttt{pair\_count}. The outputs of the two functions enter \texttt{spat\_entropy} as arguments \texttt{adj.list} and \texttt{shannZ} respectively. The output of \texttt{spat\_entropy} is a list of estimates of all quantities of Section \ref{sec:nostra}: \texttt{mut.global}, global spatial mutual information, \texttt{res.global}, global residual entropy, \texttt{shannZ}, Shannon's entropy of $Z$, \texttt{mut.local}, partial information terms, \texttt{res.local} partial residual entropies, \texttt{pwk}, spatial weights for each distance range, \texttt{pzr.marg}, the relative frequencies of $Z$, \texttt{pzr.cond}, a $K$ dimensional list with the relative frequencies of $Z$ for each distance range, \texttt{Q}, the total number of pairs and \texttt{Qk}, the number of pairs for each distance range.

In the following, we help the user through practical implementation as well as interpretation of the results.

\subsubsection{Areal data}

The workflow for computing Altieri's entropy for binary lattice data, using Bologna data, is
\begin{itemize}
\item compute Shannon's entropy of $Z$\\

\noindent\texttt{R> bo.shZ=shannonZ(boData)}\\

The function returns the pair frequencies table and the benchmark value for Shannon's entropy. Since the total number of possible pairs in the dataset is huge, this function may require a few minutes
\item build the list of adjacency matrices by choosing the distance breaks according to the case study and the observation window's size\\

\noindent\texttt{R> bo.maxdist=sqrt(diff(bo.win\$xrange)\textasciicircum2+diff(bo.win\$yrange)\textasciicircum2)\\
R> bo.distbreaks=c(0,2,4,10,bo.maxdist)\\
R> bo.adjlist=adj\_list(bo.dmat, bo.distbreaks)}\\

\item compute Altieri's entropy (which may take a few minutes on a standard laptop)\\

\noindent\texttt{R> bo.altieri=spat\_entropy(boData, bo.adjlist, bo.shZ)}\\

An extract of the output is\\

\noindent\texttt{\$mut.global\\
{[}1{]} 0.005891553}\\

\noindent\texttt{\$res.global\\
{[}1{]} 1.033506}\\

\noindent\texttt{\$shannZ\\
{[}1{]} 1.039398}\\

\noindent\texttt{\$mut.local\\
{[}1{]} 0.2365274495 0.1088072872 0.0319127643 0.0006132307}\\

\noindent\texttt{\$pwk\\
{[}1{]} 0.004642497 0.013301961 0.087665786 0.894389756}\\
\end{itemize}
Bologna's dataset can be used for assessing urban heterogeneity; in this context, a compact city represents the desirable situation, where the $X$ outcomes are highly positively correlated. In such case, spatial mutual information should be high, because urban areas generally have urban neighbours, while non-urban areas have non-urban neighbours; space plays a relevant role in determining the entropy of $Z$. In this example, the entropy \texttt{shannZ} is $H(Z)=1.039$ and its two components are \texttt{mut.global} $MI(Z,W)=0.006$ and \texttt{res.global} $H(Z)_W=1.033$. The low value of $MI(Z,W)$ is due to the low value of its components \texttt{mut.local} $PI(Z|w_k)$ at large ranges, which receive high weights \texttt{pwk} and influence the sum heavily: see the last partial values \texttt{mut.local} and the last weights \texttt{pwk}. Each partial information term measures the degree of association (compactness) in the city pattern at each distance range. The focus is on short distance ranges, where the difference between a compact city and a dispersed one is more evident: see the first \texttt{mut.local} values. By exploring these terms, an indication of the degree of dispersion can be provided. 

\begin{figure}
\centering
\includegraphics[width=.4\textwidth]{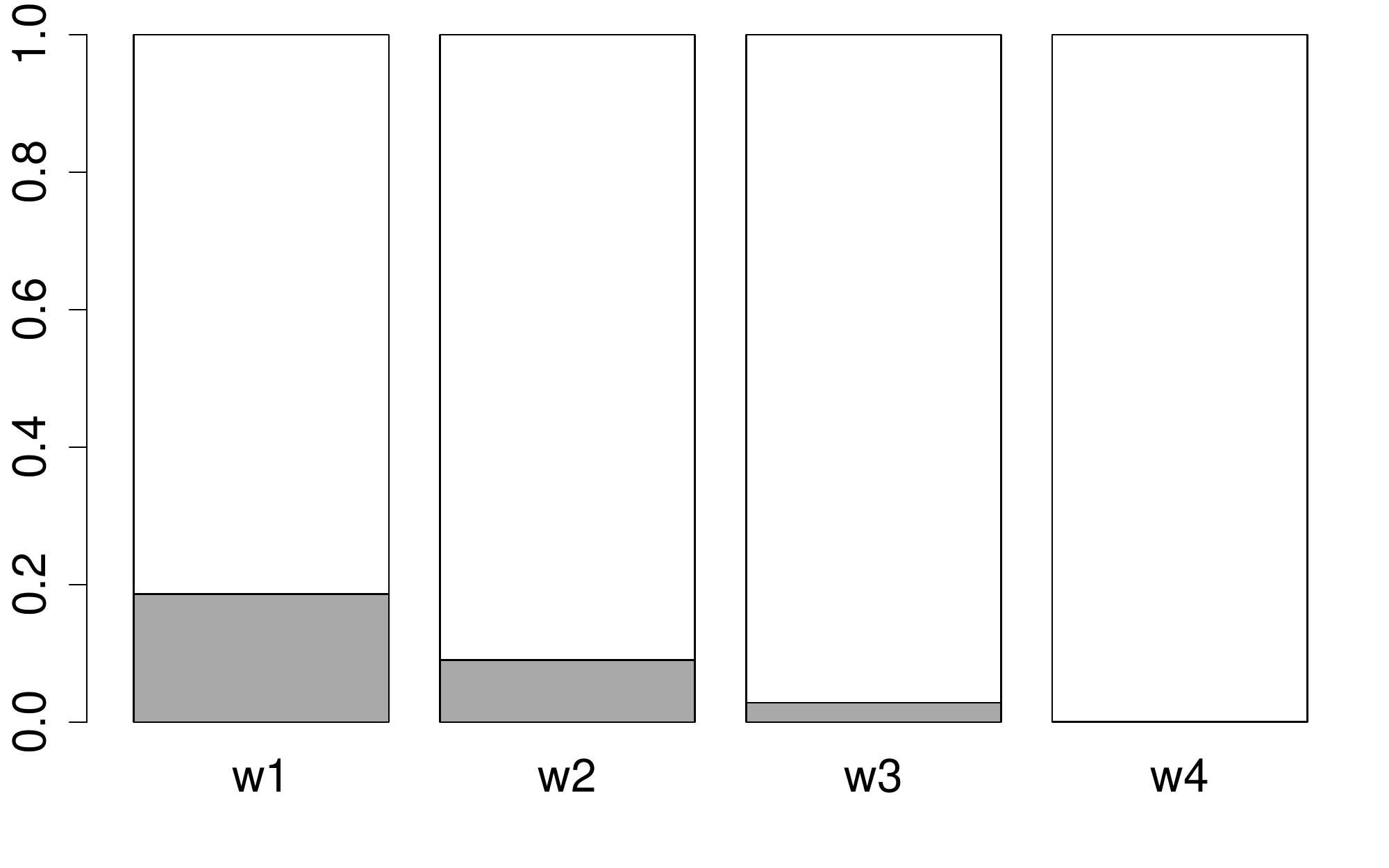}
\caption{\label{fig:bo_barre} Altieri's entropy for Bologna data: partial information (grey areas) and partial residual entropies (white areas) in proportional terms for each distance range.}
\end{figure}

Many plots can be produced for delivering the results. One example is shown in Figure \ref{fig:bo_barre}:  at each distance class, the sum $PI(Z|w_k)+H(Z|w_k)$ is set to 1, so that the contribution of space may be appreciated in proportional terms and is comparable. It is immediate to see that space explains one fifth of the data entropy at short distances, with a gradual decrease moving to large distance classes.\\

\noindent\texttt{R> local.sum=bo.altieri\$res.local+bo.altieri\$mut.local\\
R> barplot(height=rbind(bo.altieri\$mut.local/local.sum, \\
+  bo.altieri\$res.local/local.sum), beside=F,\\
+  col=c("darkgray", "white"),   names.arg=c("w1","w2","w3","w4"))}\\

\subsubsection{Point data}

The workflow for rainforest tree data is very similar, with the attention of using the vector of point marks as \texttt{data} and of tuning the distance breaks.\\

\noindent\texttt{R> tree.shZ=shannonZ(marks(treeData), missing.cat=NULL)\\
R> tree.maxdist=sqrt(diff(treeData\$win\$xrange)\textasciicircum2+diff(treeData\$win\$yrange)\textasciicircum2)\\
R> tree.distbreaks=c(0,2,4,10,tree.maxdist)\\
R> tree.adjlist=adj\_list(tree.dmat, tree.distbreaks)\\
R> tree.altieri=spat\_entropy(marks(treeData), tree.adjlist, tree.shZ)}\\

Useful guidelines for the interpretation of the results for rainforest tree data may be found in \cite{nostro}.

\section{Summary and discussion} \label{sec:summary}

In this paper, we introduce the new package \texttt{SpatEntropy}. Its version \texttt{0.1.0} contains 19 user-level functions which allow to implement spatial entropy measures. Central Sections \ref{sec:batty}, \ref{sec:oneill} and \ref{sec:nostra} separately introduce each of the main approaches to face spatial heterogeneity, and offer guidelines for choosing the most appropriate framework to measure spatial entropy, also according to the type of data. 

When the hetoregeneity of the spatial distribution of a population needs to be evaluated according to a territory partitioned into sub-areas, Batty's entropy of Section \ref{sec:batty} should be used, or its development due to Karlstr\"{o}m and Ceccato which includes a neighbourhood system. This approach has some disadvantages: a categorical variable $X$  with $I>2$ outcomes cannot be used for (\ref{eq:spaten}) and (\ref{eq:karl}), since only one category enters the measure. In other words, computations have to be conducted for each specific category of $X$, thus $I$ different entropies are computed, but no way is proposed to synthesize them into a single spatial entropy measure for $X$. Moreover, these separate conclusions are heavily affected by the choice of the area partition.

For computing the entropy of a spatially correlated categorical variable, if data are areal and the focus is on the heterogeneity of contiguous realizations, then O'Neill's entropy or the closely related contagion indices of Section \ref{sec:oneill} should be employed. Starting from a specific distance wider that contiguity and/or point data, O'Neill's extension to Leibovici's entropy of Section \ref{sec:leib} can be used. The limit of these measures is that they only provide partial results. Indeed, O'Neill's entropy only uses information about adjacent couples, and ignores the rest. Leibovici's entropy works on the same principle, extending to a general distance $d$. Thus, if $d$ is small, a great part of the spatial information is not considered; conversely, if $d$ is large, the result is aggregate and excludes any possibility to explore the contribution of space in detail.

Lastly, if one wants to take a complete approach, which considers not only a single distance but all possible distance ranges to evaluate the overall influence of the spatial configuration in computing the data entropy, and which allows a flexible decomposition, then the recent Altieri's approach of Section \ref{sec:nostra} is the most appropriate choice. The additive terms (\ref{eq:partialterm_mut}) and (\ref{eq:residZW_loc}), together with their sums (\ref{eq:PI_add}) and (\ref{eq:residZWadd}), constitute the set of spatial entropy measures. The approach is able to: maintain the information about the categories of $X$; consider different distance ranges simultaneously, by including an additional study variable representing space to enjoy the properties of bivariate entropy measures; quantify the overall role of space; be additive and decomposable. Therefore, spatial mutual information has in this case theoretical support to be considered the most reliable method for measuring data heterogeneity; it is also easily interpretable.

The package \texttt{SpatEntropy} works for areal and point data presenting a number of categories $I\ge 2$. It includes all necessary functions for extracting the spatial entropy of the data from scratch: the practical parts of Sections \ref{sec:batty}, \ref{sec:oneill} and \ref{sec:nostra} give step-by-step details.

User feedback is a fundamental part of the package development process. We welcome feedback and suggestions from all users.

\section*{Acknowledgments}

This work is developed under the PRIN2015 supported project 'Environmental processes and human activities: capturing their interactions via statistical methods (EPHASTAT)' [grant number 20154X8K23] funded by MIUR (Italian Ministry of Education, University and Scientific Research).

\bibliographystyle{chicago}
\bibliography{bibdatabase_entropy}

\end{document}